\documentclass[a4paper,11pt]{article}
\usepackage[paper=a4paper,left=9mm,right=9mm,top=20mm,bottom=25mm]{geometry} 
\pdfoutput=1

 \usepackage{setspace}
\usepackage{authblk}

\usepackage{fullpage}
\usepackage{parskip}
\usepackage{titlesec}
\usepackage[section]{placeins}
\usepackage[dvipsnames, table]{xcolor}
\usepackage{enumitem}

\usepackage{breakcites}
\usepackage{lineno}
\usepackage{hyphenat}
\usepackage[all]{nowidow}

\usepackage{centernot} % für Befehl \centernot
\usepackage{lscape}

\PassOptionsToPackage{hyphens}{url}
\usepackage{hyperref} 
\hypersetup{colorlinks = true}

\hypersetup{
  linkcolor  = RedViolet,
  citecolor  = Blue!90!black,
  urlcolor   = Aquamarine!80!black,
  colorlinks = true
}

\usepackage{etoolbox}
\usepackage{natbib}
\renewenvironment{abstract}
  {{\bfseries\noindent{\abstractname}\par\nobreak}\footnotesize}
  {\bigskip}
\titlespacing{\section}{0pt}{*3}{*1}
\titlespacing{\subsection}{0pt}{*2}{*0.5}
\titlespacing{\subsubsection}{0pt}{*1.5}{0pt}

\usepackage{graphicx}
\usepackage[space]{grffile}
\usepackage{latexsym}
\usepackage{textcomp}
\usepackage{tabulary}
\usepackage{booktabs,array,multirow}
\usepackage{amsfonts,amsmath,amssymb}
\usepackage{subcaption}
\usepackage{booktabs} % für toprule, midrule und bottomrule 
\usepackage{tabularx}
\usepackage{amssymb}
\usepackage{tabularx}
\usepackage{pifont}
\usepackage{caption} % to center a caption
\usepackage{hyperref}
\usepackage{comment}
\providecommand\citet{\cite}
\providecommand\citep{\cite}

\newif\iflatexml\latexmlfalse

\usepackage{comment}
\usepackage{tabularx}
\usepackage{rotating}
\usepackage{wrapfig}
\usepackage{threeparttable}

\AtBeginDocument{\DeclareGraphicsExtensions{.pdf,.PDF,.eps,.EPS,.png,.PNG,.tif,.TIF,.jpg,.JPG,.jpeg,.JPEG}}

\usepackage[utf8]{inputenc}
\usepackage[english]{babel}

\usepackage{orcidlink}
\newcommand{\beginsupplement}{
        \setcounter{section}{0} 
        \renewcommand{\thesection}{S\arabic{section}}

        \setcounter{table}{0}
        \renewcommand{\thetable}{S\arabic{table}}
        \setcounter{figure}{0}
        \renewcommand{\thefigure}{S\arabic{figure}}
     }

     \newcolumntype{Y}[1]{>{\centering\arraybackslash}p{#1}}

\newcommand{\newcrossmark}{\scalebox{1}[1]{$\times$}}

\begin{document}
\title{Rethinking the handling of method failure in comparison studies}

\def\correspondingauthor{\footnote{Corresponding author, e-mail: \href{mailto:milena.wuensch@ibe.med.uni-muenchen.de}{milena.wuensch@ibe.med.uni-muenchen.de}, Institute for Medical Information Processing, Biometry, and Epidemiology, LMU Munich, Marchioninistr. 15, D-81377, Munich, Germany}}
\author[1,2]{Milena W{\"u}nsch \correspondingauthor{} \orcidlink{0009-0001-1982-9260}}
\author[1,2]{Moritz Herrmann\orcidlink{000-0002-4893-5812}}

\author[3]{Elisa Noltenius}
\author[3]{Mattia Mohr}
\author[4]{Tim~P.~Morris \orcidlink{0000-0001-5850-3610}}
\author[1,2]{Anne-Laure Boulesteix \orcidlink{0000-0002-2729-0947}}
\affil[1]{Institute for Medical Information Processing, Biometry, and Epidemiology, Faculty of Medicine, LMU Munich (Germany)}
\affil[2]{Munich Center for Machine Learning (MCML), Munich (Germany)}
\affil[3]{Department of Statistics, LMU Munich, Munich (Germany)}
\affil[4]{MRC Clinical Trials Unit, UCL, London (UK)}
\vspace{-1em}
  \date{\today}
\maketitle

%\author[1,2,$\ast$]{Milena Wünsch \ORCID{0009-0001-1982-9260}}
%\author[1,2]{Christina Sauer \ORCID{0000-0003-2425-7858}}
%\author[1]{Patrick Callahan \ORCID{0000-0003-1769-7580}}
%\author[3]{Ludwig Christian Hinske}
%\ORCID{0000-0001-7273-5899}
%\author[1,2]{Anne-Laure Boulesteix}
%\ORCID{0000-0002-2729-0947}

\begin{abstract}
Comparison studies in methodological research are intended to compare methods in an evidence-based manner to help data analysts select a suitable method for their application. To provide trustworthy evidence, they must be carefully designed, implemented, and reported, especially given the many decisions made in planning and running. A common challenge in comparison studies is to handle the ``failure'' of one or more methods to produce a result for some (real or simulated) data sets, such that their performances cannot be measured in those instances. Despite an increasing emphasis on this topic in recent literature (focusing on non-convergence as a common manifestation), there is little guidance on proper handling and interpretation, and reporting of the chosen approach is often neglected. This paper aims to fill this gap and offers practical guidance on handling method failure in comparison studies. After exploring common handlings across various published comparison studies from classical statistics and predictive modeling, we show that the popular approaches of discarding data sets yielding failure (either for all or the failing methods only) and imputing are inappropriate in most cases. We then recommend a different perspective on method failure---viewing it as the result of a complex interplay of several factors rather than just its manifestation. Building on this, we provide recommendations on more adequate handlings of method failure derived from realistic considerations. In particular, we propose considering fallback strategies that directly reflect the behavior of real-world users. Finally, we illustrate our recommendations and the dangers of inadequate handling of method failure through two exemplary comparison studies.

\end{abstract}

\sloppy

\section{Introduction}

In methodological research, comparison studies aim to compare the performance of methods to provide empirical evidence on their behavior and help data analysts choose appropriate methods for their setting. To be reliable, they require careful planning of the design, execution, and reporting of the results. These issues have gained increasing attention in recent years, both in the context of simulation studies based on artificially generated data \citep{morris2019using, pawel2024pitfalls, siepeprereg} and in the context of real data-based benchmark studies \citep{boulesteix2017towards,friedrich2024role}. 

In this work, we focus on a particular aspect of the design of comparison studies that affects both simulation and benchmark studies, namely ``method failure.'' Method failure occurs when a method under investigation fails to produce an output for a (real or simulated) data set considered in the comparison study. 
The issue of method failure in comparison studies has begun to attract attention in recent years \citep{morris2019using, pawel2024pitfalls, siepeprereg}, often focusing on non-convergence. However, there is little guidance on how to handle method failure appropriately, whether it occurs as non-convergence or in another form. 
In particular, published reviews have shown that method failure goes unreported in the majority of cases. For instance, in a systematic review of $42$ published simulation studies on the analysis of complex longitudinal patient-reported outcomes data, less than half \textit{``acknowledge that model non-convergence might occur''}  \citep{hinds2018systematic}. Even more notably, only $12$ of $85$ applicable research articles reviewed by \citet{morris2019using} report convergence as a performance measure. This contrasts with our perception---both through our own experiences and informal discussions with colleagues---that the problem of method failure is highly prevalent. In particular, signs of method failure are often traceable in the code provided for reproducibility, even when not reported in the manuscript. This is often clear through the use of explicit error-handling mechanisms (e.g., functions \texttt{try()} or \texttt{tryCatch()} in \texttt{R} or \texttt{try-except} constructs in \texttt{Python}).

To illustrate the difficulty of handling method failure adequately, imagine a researcher conducting a comparison study in statistics or predictive modeling to compare the performance of three methods based on four simulation/benchmark data sets. In the context of classical statistics, they likely focus on comparing the methods' performances (e.g., bias) in \textit{``estimating one or more population quantities''} \citep{morris2019using}, sometimes called ``estimand(s),'' and base the analysis on data simulated using a certain data-generating mechanism (DGM). In predictive modeling, they are likely to compare the methods' predictive performances (e.g., accuracy), that is, their ability to make predictions on a new set of observations. Here, the analysis is typically performed on real benchmark data sets. For simplicity, we refer to both settings as ``simulation studies'' and ``benchmark studies,'' respectively (though admitting that there may be intermediate situations, e.g., simulation studies investigating predictive performance on artificial data). 

During the experiments, the researcher encounters the failure of some methods in one or more of the following ways: (i) they receive an error message or output of \textit{NA} (``not applicable'') or \textit{NaN} (``not a number''), implying that a certain calculation could not be completed (e.g., non-convergence), (ii) their system crashes, or (iii) the experiment runs for an excessively long time, forcing the researcher to abort.

While the majority of real published comparison studies do not report method failure, there are positive examples from classical statistics and predictive modeling in which the occurrence (and oftentimes also the applied handling) of method failure is described. For instance, \citet{vansmeden2016_10epv} investigate why different simulation studies offer conflicting minimal ``events per variable (EPV)'' recommendations for logistic regression. All three studies featured, exploring the effect of EPV on performance measures such as bias and coverage, encounter separated data sets (i.e., outcome variable is perfectly predicted by one or a combination of multiple covariates), which lead to \textit{convergence issues} of the maximum likelihood process. That is, convergence is not achieved within the pre-specified number of iterations, or the process converges to a point that is not the maximum likelihood estimate \citep{vansmeden2016_10epv}. In a study from the statistical learning field, \citet{gijsbers2024amlb} conduct a comparison study of nine automated machine learning methods (``AutoML frameworks''). They encounter and report method failure in the forms of (i) \textit{exceeding available memory} (or other memory-related issues), (ii) \textit{exceeding time limits}, (iii) \textit{failure related to data set characteristics} (such as highly imbalanced data), and (iv) failure caused by \textit{errors in the implementation} of the AutoML method \citep{gijsbers2024amlb}. 

\begin{table}[h]
\small
\caption{Overview table of (a) a fictive simulation study across four repetitions (rows) and (b) a fictive benchmark study across four benchmark data sets (rows) for three methods (columns). Method failure results in undefined values for the affected combination of data set and method, marked by ``NA.''} 
\label{tab:three_tables}
\begin{subtable}{0.48\textwidth}
\centering
\caption{Fictive simulation study: overview of estimates, with the true value being equal to $4$. Method failure first leads to an undefined estimates value for the given method--data set combination, complicating the derivation of bias for that method. }\label{tab:overview_missings_simulation}
\vspace{0.2cm}
\renewcommand{\arraystretch}{1.2} % increase line spacing
    \begin{tabular}{c|cccc}
    \hline
        Repetition &  Method 1 & Method 2 & Method 3\\
        \hline
         1  & 3.89 & 4.23 &  4.08\\
         2  & 3.78 & 4.13  &  4.11\\
         3  & \textbf{NA} & 3.69 &  4.23\\
         4  & 3.75 & 4.24 & \textbf{NA}\\
         \hline
         Bias & \textbf{?} & 0.07 &  \textbf{?}\\
    \end{tabular}
\end{subtable}
\hfil
\begin{subtable}{0.48\textwidth}
\centering
   \caption{Fictive benchmark study: overview of estimated accuracies across all method--data set combinations. Method failure directly leads to an undefined performance (i.e., accuracy) value for the given method--data set combination. \\}\label{tab:overview_missings_benchmark}
   \vspace{0.2cm}
   \renewcommand{\arraystretch}{1.2} % increase line spacing
    \begin{tabular}{c|cccc}
    \hline
        Benchm. data set &  Method 1 & Method 2 & Method 3\\
        \hline
         1  & 0.85 & 0.88 &  0.87\\
         2  & 0.9 & 0.91  &  0.86\\
         3  & \textbf{NA} & 0.80 &  0.82\\
         4  & 0.78 & 0.76 & \textbf{NA}\\
         \hline
         Av. accuracy & \textbf{?} & 0.84 &  \textbf{?}\\
    \end{tabular}
\end{subtable}
\end{table}

Regardless of the form in which a researcher encounters method failure, it leads to NA for the associated method-data set combination where otherwise an estimate (for simulation studies) or a performance value (for benchmark studies) is expected. Table~\ref{tab:three_tables} depicts a corresponding summary of the results of the hypothetical researcher's study across all simulation repetitions/benchmark data sets and methods. In simulation studies, the NAs complicate the derivation of performance (e.g., bias as the mean deviation of the estimates from the true value across all repetitions). In benchmark studies, on the other hand, they hinder the aggregation of performance values into an average performance (e.g., average estimated accuracy as the mean fraction of correct predictions across all benchmark data sets). In both cases, method failure complicates the assessment and especially the comparison of performance of the methods under investigation such that the researcher finds themself wondering how to proceed adequately. 

The above-shown examples of published comparison studies alone already illustrate different viewpoints on method failure and correspondingly different ways in which it is handled in practice. \citet{vansmeden2016_10epv} observe that separated \textit{data sets are commonly removed} from the analysis. This, however, affects bias, mean squared error (MSE), and confidence interval width, even if the frequency of separated data sets is small \citep{vansmeden2016_10epv}. \citet{gijsbers2024amlb} argue against discarding data by describing that method failure is typically correlated with data set characteristics. Instead, they \textit{impute} by the performance of a constant predictor. \\
Examining additional published studies, which we do throughout this paper, shows that imputation and discarding data sets associated with failure are among the most common practices.
However, it also becomes clear that they are only two of many handlings applied in practice and that others, such as modifying the affected methods' implementations and changing software, are common as well---an indicator that proper guidance is commonly lacking. Our paper aims to fill this gap and provides detailed instructions on how to handle method failure. In particular, we illustrate why usual handlings of method failure derived from treating the resulting NAs as regular ``missing data,'' and especially imputation and discarding data, are inappropriate in most settings (leading us to refer to ``undefined'' rather than ``missing'' performance values in instances of method failure). As an alternative, we promote considering method failure from a different viewpoint, namely as the result of a complex and individual interplay of multiple factors---which we use to derive alternative and more appropriate measures that can be applied when encountering method failure. \\
Our paper targets both the planning and the execution stage of a comparison study. While many instances of method failure can only be directly observed---and therefore investigated---after the study is executed, certain considerations should be made while planning to anticipate and ideally reduce the potential for failure. \\
The rest of this paper is structured as follows. In Section~\ref{sec2}, we give an overview of the status quo by reviewing and summarizing the different manifestations and handlings of method failure reported in published comparison studies, followed by a detailed discussion of why discarding data or imputing values is usually inadequate when encountering method failure. Building on this, Section~\ref{sec3} is a compilation of alternative recommendations for handling method failure, which differentiates between different scopes of the study and research stages in which it may take place. In particular, we promote using fallback strategies when encountering method failure, which enables the aggregation of performance across data sets even when method failure occurs, and on top of that, reflects the behavior of real method users across many settings. Considering possible time constraints, we also discuss ``slimmed-down'' ways in which authors can implement part of our recommendations if unable to follow them completely. Finally, in Section~\ref{sec4}, we demonstrate our recommendations using two fictive comparison studies that illustrate the complexity of the discussed issues.

\section{Method failure in published comparison studies}\label{sec2}
In Section~\ref{sec:state-of-the-art}, we summarize common manifestations and handlings of method failure reported in the literature. Since some authors focus on how method failure is handled rather than explaining how it occurred, the corresponding studies are discussed only in Section~\ref{sec:applied_handlings}. An overview of all study contexts, manifestations, and handlings of method failure are provided in Table~\ref{tab:overview_literatureexamples}.

\subsection{Status quo: occurrence and handling of method failure}\label{sec:state-of-the-art}

\begin{tiny}
\begin{sidewaystable}
\resizebox{\textwidth}{!}{
\begin{threeparttable}
    
\caption{Selection of comparison studies from statistics and predictive modeling that report the occurrence and handling of method failure.} \label{tab:overview_literatureexamples}
\scriptsize

\centering

     \begin{tabular}{|p{3cm}|p{12cm}!{\vrule width 1.2pt}*{2}{p{0.4cm}|}l!{\vrule width 1.2pt}*{7}{p{0.3cm}|}}

    \hline
   \multicolumn{2}{|c!{\vrule width 1.2pt}}{Comparison study } &\multicolumn{3}{c!{\vrule width 1.2pt}}{Manifestations} & \multicolumn{7}{c|}{Handlings}\\ 
   \hline
   \multicolumn{2}{|l !{\vrule width 1.2pt}}{} &\multicolumn{3}{c!{\vrule width 1.2pt}}{} & \multicolumn{3}{c}{Resolve }& \multicolumn{4}{|c|}{Summarize despite }\\
      \multicolumn{2}{|l !{\vrule width 1.2pt}}{} &\multicolumn{3}{c!{\vrule width 1.2pt}}{} & \multicolumn{3}{c}{method failure}& \multicolumn{4}{|c|}{ method failure}\\
   \hline

Author & Context of the study &\begin{turn}{90}
Non-convergence (and others) \end{turn} & \begin{turn}{90}Runtime issues \end{turn}& \begin{turn}{90}Memory issues \end{turn}& \begin{turn}{90}Remove method \end{turn}& \begin{turn}{90}Modify implementation \end{turn}& \begin{turn}{90} Change software \end{turn} & 
\begin{turn}{90}Discard data for failing methods \end{turn} & \begin{turn}{90} Discard data for all methods \end{turn}&\begin{turn}{90} Impute \end{turn} & \begin{turn}{90} Include performance measure \end{turn} \\
\hline

\rowcolor{gray!25}    
\citet{vansmeden2016_10epv}\tnote{a}  & 
Simulation study to investigate why different published simulation studies offer conflicting minimal ``events per variable (EPV)'' recommendations for logistic regression. Features three published simulation studies exploring the effect of EPV on performance measures. 
&\newcrossmark& & && &  &  & \newcrossmark &  &  \\
\rowcolor{gray!7}
\citet{gijsbers2024amlb} &  Comparison study of nine automated machine learning (``AutoML'') methods. & \newcrossmark& \newcrossmark& \newcrossmark &  &  &  &  &  & \newcrossmark & \\
\rowcolor{gray!25}
\citet{zapf2024meta}  & Simulation study to compare three frequentist methods for the meta-analysis of diagnostic accuracy studies regarding sensitivity and specificity. & \newcrossmark& &  &  &  & & \newcrossmark &  &  &  \newcrossmark\\
\rowcolor{gray!7}
\citet{ruxton2013review} & Simulation study on the empirical coverage of eight methods for calculating confidence intervals for the odds ratio for binary outcome and exposure. & \newcrossmark&& & \newcrossmark&  &  & &  &  &   \\
\rowcolor{gray!25}
\citet{hornung2024prediction} & Benchmark study using thirteen multi-omics data sets to assess the performance of seven prediction methods for multi-omics data that can handle block-wise missingness by artificially inducing specific patterns of block-wise missing values into initially complete data.&\newcrossmark & & \newcrossmark &  &  & &  & \newcrossmark &  &  \\
\rowcolor{gray!7}
\citet{masaoud2010simulation}& Two simulation studies to compare the performance of seven marginal and random effects estimation methods for binary longitudinal data (e.g., GEE, Bayesian Markov Chain Monte Carlo). & \newcrossmark& &  & \newcrossmark & \newcrossmark & \newcrossmark &  &  & &  \\
\rowcolor{gray!25}
\citet{crowther14}  & Simulation study to assess the performance of a multilevel mixed effects parametric survival model using adaptive and nonadaptive Gauss-Hermite quadrature. &\newcrossmark & &  &  & \newcrossmark& &  &  & &  \\
\rowcolor{gray!7}
\citet{dunias2024comparison} & Simulation study to compare the performance of five hyperparameter tuning procedures for clinical prediction models built upon statistical learning methods.&\newcrossmark& &  &  &  & &  &  & \newcrossmark  &  \\
\rowcolor{gray!25}
\citet{fernandez2014we} & Benchmark study of 179 supervised classification methods across 121 benchmark data sets. &\newcrossmark& & \newcrossmark &  &  & & \newcrossmark &  & \newcrossmark  &  \\
\rowcolor{gray!7}
\citet{herrmann2021large} & Benchmark study on the predictive performance of thirteen survival time prediction methods based on eighteen high-dimensional multi-omics data.&*& \newcrossmark & * &  \newcrossmark\tnote{b}&  & \newcrossmark&  &  & \newcrossmark  &  \\
\rowcolor{gray!25}
\citet{bischl2013benchmarking} &Comparison study of eleven classification methods on 49 real benchmark and simulated data sets. &*&*&*  & &  & &  &  & \newcrossmark  &  \\
\rowcolor{gray!7}
\citet{niessl2022over}\tnote{c} & Meta-study to investigate the effect of different imputation procedures on the performance assessment and ranking of the methods. Considered alternative procedures include imputation by the worst possible value and by a value that weighs the method's overall proportion of method failure against its performance in the remaining folds. &*&*&* &  &  & & &  & \newcrossmark  &  \\

 \hline
    \end{tabular}
        \begin{tablenotes}
       \item [a] \citet{vansmeden2016_10epv} \textit{observe} in published comparison studies that separated data sets (i.e., data sets leading to non-convergence) are commonly omitted from analyses.
       \item[b] \citet{herrmann2021large} apply a moderate form of data removal by restricting the analysis to the subset of smaller benchmark data sets for one method 
       \item [c] \citet{niessl2022over} perform an example benchmark study based on the results of the study by \citet{herrmann2021large} to illustrate how different ways of imputation can influence the results of a benchmark study.
       \item [*] The authors did not report details on the manifestation of method failure. \end{tablenotes}
    \end{threeparttable}
        }
    \end{sidewaystable}
    \end{tiny}
\subsubsection{Common manifestations of method failure}
\label{sec:manifestations}

\textbf{Non-convergence and other calculation issues} \ Authors often encounter the inability of a method to perform a calculation necessary for prediction or estimation, either implied through an error, a warning, or a meaningless output. For methods employing an iterative procedure (e.g., an iterative maximum-likelihood procedure), this is called ``non-convergence,'' i.e., the method fails to produce a valid output within the pre-specified number of iterations. 

\textit{Examples} \ \citet{zapf2024meta} encounter non-convergence for two of the three methods under investigation. One returns an error, while the other outputs the current, non-converged estimate in these instances \citep{zapf2024meta}. \citet{masaoud2010simulation} encounter instances of \textit{``non-convergence or non-sensible estimates''} for some method-data set combinations. In the study by \citet{crowther14} the methods are based on numerical integration and encounter estimation difficulties (non-convergence) with \textit{``challenging data sets''} \citep{crowther14}. 
\citet{hornung2024prediction} report instances of method failure when methods are \textit{``not applicable''} to certain differences between training and test sets regarding the patterns of block-wise missingness. Additionally, they report calculation issues for one method explicitly disallowing certain structures in the training data set, resulting in failure. 
Calculation issues are also reported in the study by \citet{ruxton2013review}, where some odds ratio estimation methods fail due to division by $0$ when the $2\times 2$ contingency table contains one or more zero entries. 
\citet{fernandez2014we} report several data set characteristics associated with issues in required calculations, including \textit{``collinearity of data, singular covariance matrices, and equal [predictor values] for all the training [observations], [\dots] discrete predictor variables, classes with low populations, or too few classes''} \citep{fernandez2014we}. Additionally, a group of methods in their study requires a certain minimum number of observations per outcome class and is therefore not applicable to particularly small data sets.
Another case of calculation issues can be found in the simulation study conducted by \citet{dunias2024comparison}. For some data sets, the statistical learning method LASSO does not select any predictor variable, leading to a constant predictor and preventing the assessment of performance regarding calibration slope. \\
\textbf{Memory issues} \ Even before exceeding possible time budgets, some methods crash due to memory issues. That is, the given method consumes more memory than is available. \\
\textit{Examples} \ \citet{gijsbers2024amlb} report memory constraint violations, segmentation faults, and Java server crashes as instances of method failure related to memory. \citet{hornung2024prediction} also encounter memory issues, which they attribute to a specific missingness pattern induced into the training data. 
\citet{fernandez2014we} encounter memory issues in combination with large data sets. 

\textbf{Runtime issues} \ Another form of method failure often reported in comparison studies is excessive runtime. Some studies set fixed time budgets, so exceeding them is considered method failure. Others without fixed budgets encounter method failure when experiments for single methods run for an excessive amount of time (e.g., days), forcing the researcher to abort. Unlike non-convergence and memory issues, method failure is \textit{actively declared} by the researchers when runtime issues occur, as without these (naturally necessary) limitations, a method might eventually generate an output at some point. Resulting implications are discussed in Section~\ref{sec:runtime_issues}.\\
\textit{Examples} \ While \citet{gijsbers2024amlb} encounter the excess of predefined time budgets for model training, \citet{herrmann2021large} face runtime issues without fixed budgets. They encounter \textit{``computations lasting several days for one single model fit for large data sets''} \citep{herrmann2021large} for one method. 

\subsubsection{Common approaches to handling method failure}\label{sec:applied_handlings}

Based on the published comparison studies, the handling of method failure can be commonly divided into two general approaches. Note that a literature review by \citet{pawel2024handling}, yielding a structured investigation of the frequencies of handling approaches from published simulation studies, results in four main handlings, which are among those we have commonly observed. 

\textbf{Resolving method failure} \ 
Some authors resolve method failure completely so that no instances remain. A straightforward way observable in the literature is to completely \textit{remove the affected method from the analysis}. For instance, \citet{ruxton2013review} exclude all methods that fail with zeros in the $2\times 2$ contingency table. \citet{masaoud2010simulation} use method removal as a last resort in their sequence of handlings when previous measures (see below) have been unsuccessful. This illustrates well that oftentimes, different handlings are applied within the same study, sometimes switching between approaches or applying them sequentially until successful.  Note that a more moderate form of method removal can be observed in the benchmark study by \citet{herrmann2021large}. For the method encountering runtime issues, they restrict the analysis to the subset of smaller benchmark data sets and report the corresponding results separately from the remaining methods.  

An alternative to removing methods is to \textit{modify their implementation} such that an output can be successfully produced. For instance, in the study by \citet {masaoud2010simulation}, \textit{``different optimization techniques were tried''} \citep{masaoud2010simulation} for one method when the default procedure resulted in failure. \citet{crowther14} react to method failure by successively increasing the number of quadrature points until convergence is reached. \\
Lastly, \textit{changing software} for the method affected by method failure can also be observed in the literature. Switching the operating system can be observed in the study by \citet{herrmann2021large}, who for one method encounter \textit{``a fatal error in R under Windows, but not using the Linux distribution Ubuntu 14.04''} \citep{herrmann2021large}. \citet{masaoud2010simulation}, while remaining in the same operating system, generate the required output for one method in \texttt{SAS} when it fails in \texttt{R}. 

\textbf{Summarizing performance despite method failure}
Instead of resolving method failure, some authors summarize the methods' performances despite existing failures. A widespread approach is to discard data sets associated with failure. The first version of this approach, consisting of \textit{discarding data sets for the failing methods only}, is followed by \citet{zapf2024meta} for the method affected by errors due to non-convergence. \citet{fernandez2014we}, comparing the methods' performances based on two performance measures, discard the associated data sets for the methods affected by method failure in the comparison based on average accuracy. The second version of discarding data, namely \textit{discarding the corresponding data sets for all methods}, is applied by \citet{hornung2024prediction}. Particularly, they argue against discarding the data sets for the failing methods only by saying that it would not yield a fair comparison since predictive performance is generally associated with the induced pattern of block-wise missingness \citep{hornung2024prediction}. \\
Instead of discarding data, it is also common to \textit{impute performance values} for the instances of method failure. Note that imputation yields a variety of options. \citet{fernandez2014we}, for instance, impute by the mean accuracy of all remaining methods on the given data set for their second performance measure (Friedman ranking) when a method fails as it requires the same number of performance values across all methods. On the other hand, recall \citet{gijsbers2024amlb} and their comparison study of AutoML methods in which they impute with a constant predictor when a method fails. They choose this \textit{``very penalizing imputation strategy}'' \citep{gijsbers2024amlb}, which can be viewed as a form of imputation by the worst possible value, with a similar reasoning to \citet{hornung2024prediction}. That is, failure may be correlated with data set characteristics such that discarding data sets for the failing methods or imputing by the average performance of the given method on the remaining data sets may unfairly favor the failing methods \citep{gijsbers2024amlb}. Note that \citet{dunias2024comparison} also impute by a constant predictor when a statistical learning method fails. \\
An alternative way of imputing is employed by \citet{bischl2013benchmarking}, who impute performance values based on the proportion of repetitions in which the affected method fails. Below $20\%$, they impute by sampling from a normal distribution estimated from the remaining performance values of that model. Above $20\%$, on the other hand, they impute by the worst possible value with the reasoning that the method's behavior is \textit{``too unreliable for the current data set''} \citep{bischl2013benchmarking}. A similar imputation strategy is employed \citet{herrmann2021large}. While using the same imputation rule when a method fails on more than $20\%$ of data sets, they otherwise impute by the mean performance of the method in the remaining ``successful'' data sets. 
In a subsequent project, \citet{niessl2022over} consider alternative procedures, including imputation by the worst possible value and by a value that weights the method's overall proportion of method failure against its performance in the remaining folds. They show that the approach to imputing the undefined performance values has a notable effect on the performance ranking between the methods. \\
When summarizing performance despite the failure of an existing method, some authors \textit{report the proportion of (real or simulated) data sets for which each method fails}. For instance, \citet{hornung2024prediction} list ``the frequency of data sets with missing results'' \citep{hornung2024prediction} for the methods affected by method failure. \citet{gijsbers2024amlb} provide multiple failure plots, consisting of the number of failures per method with differentiation of the different failure types, the number of failures by size of the associated data set, and boxplots of the training durations for each method with indicators of excessive runtimes. \citet{zapf2024meta} even include convergence as a performance measure, supported by a table of the number of converged simulation runs for each simulation scenario across the methods in the supplement. \\

\subsection{Why popular handlings are often inadequate}\label{sec:pitfalls_general}

\subsubsection{Why discarding data is inadequate}\label{sec:caution_discard_impute}
Currently, there is no one-size-fits-all solution to handling method failure, exemplified by the examples in Section~\ref{sec2}. This section discusses some principles to guide decisions. To illustrate ideas, we begin with two simple examples and some useful terminology. Suppose a simulation study aims to compare the performance of method~A with method~B in terms of bias. Method~A fails in $20\%$ of repetitions while method~B does not fail in any. Alternatively, suppose a benchmark study aims to compare the average accuracy of two methods. Again, method~A fails in $20\%$ of benchmark data sets while method~B does not fail in any. The hypothetical results for the two cases are given in Table~\ref{table:AB}~(a) and (b).

\begin{table}[!htb]
    \caption{Overview table of an illustrative comparison study comparing the performance of (a) two estimation methods regarding bias and (b) two prediction methods regarding average accuracy. Method A fails in $20\%$ of (a) simulation repetitions and (b) benchmark data sets.}
    \begin{subtable}{0.5\linewidth}
      \centering
        \caption{Hypothetical simulation study}
           \renewcommand{\arraystretch}{1.1} % increase line spacing
    \begin{tabular}{ccc}
        \hline
         & & Conditional   \\
        Bias & Unconditional & (repetitions where   \\
         &(all repetitions) & A does not fail) \\ \hline
        Method~A & -                    & 0.09 \\
        Method~B & 0.14 & 0.09 \\ \hline
    \end{tabular}
 \end{subtable}%
    \begin{subtable}{.5\linewidth}
        \centering
        \caption{Hypothetical benchmark study}
           \renewcommand{\arraystretch}{1.1} % increase line spacing
     \begin{tabular}{ccc}
        \hline
        & & Conditional \\
        Average & Unconditional   &  (data sets where  \\
        accuracy & (all data sets) & A does not fail) \\ \hline
        Method~A & -                    & 0.85 \\
        Method~B & 0.72 & 0.85 \\ \hline
    \end{tabular}
    \end{subtable}
\label{table:AB}
\end{table}

We contrast the terms \textit{unconditional vs.\ conditional} performance and \textit{absolute vs.\ relative} performance. \textit{Unconditional} performance simply means ``in all data~sets,'' while \textit{conditional} means ``in data~sets where this method did not fail.'' The \textit{unconditional} bias of method~B is 0.14, while the unconditional bias of method~A is not defined. The conditional bias of both methods is 0.09. Parallel descriptions apply to the hypothetical benchmark study in Table~\ref{table:AB}~(b). \textit{Absolute} performance means ``considered on its own'' and \textit{relative} performance means ``when compared with another method.'' 

These terms help to distill the problems with method failure in comparison studies. Clearly, we want to compare the methods’ unconditional performances. However, in both Tables~\ref{table:AB}~(a) and \ref{table:AB}~(b), the relative unconditional performance of methods cannot be compared because the unconditional performance of method~A is undefined. In contrast, the relative conditional performance can be quantified and compared (``discarding data sets associated with failure for all methods’’): in Table~\ref{table:AB}~(a), the two methods have the same bias conditional on non-failure of method~A; in Table~\ref{table:AB}~(b), the two methods have the same accuracy conditional on non-failure of method~A. However, the \textit{conditional} performance of method~B should be of little interest. Why should a method’s performance depend on the behavior of another method regarding failure? In Table \ref{table:AB}, method~B’s performance improves when assessed based on only those data sets where A does not fail.  This becomes even more complex as additional methods are included in the comparison, since discarding the corresponding data sets for all methods requires conditioning on the non-failure of each one.

Alternatively, one could intuitively proceed to compare the unconditional performance of method B with the conditional performance of method~A (``discard data sets for the failing methods only’’). However, comparisons should be performed vertically, or we are not comparing like with like: In the context of Table~\ref{table:AB}, this would reflect poorly on method~B, even though both perform equally well for the data sets without failure.  This is all the more problematic since method B, in this case, is actually preferable to method A, as it outputs a result for all data sets! Again, this becomes even more pronounced as additional methods are added to the comparison, since different methods typically fail on different datasets (see Section~\ref{sec:interplay_method_data}).

\subsubsection{Why imputation is inadequate}

Another superficial solution often proposed when looking at results such as those in Table~\ref{table:AB} is to interpret method failure as causing missing values and use missing data methods, such as imputation, to infer the unconditional performance of method~A. After all, results given by different methods tend to be positively correlated, so results given by method~B could inform the imputation (analogously, it may be intuitive to impute with the results of method~A on the remaining data sets as results tend to be correlated).

We view it as mistaken to regard method failure as a form of missing data and have been careful to refer to ``undefined'' rather than ``missing'' values. The missing data literature defines missing values as those that exist and were intended to be recorded but for one reason or another are \textit{``not observed''} \citep{rubin2004multiple, carpenter2023multiple}. Method failure is far closer to what is called a truncating event in the estimands literature \citep{Kurland2009,Kahan2024}, which denotes a value that does not exist. That is, when a method fails, the result we wished to record simply does not exist for that data set. Method failure should, therefore, not be viewed as obscuring the method's results on a given data set. Instead, failure is the result we get! For this reason, we should be cautious about applying missing data approaches, and particularly imputation, to method failure. Note that this also includes measures of ``weighted performance,'' which weight the performance of a method according to the proportion of data sets with failure for that method. Using the correspondence between weighting and imputation approaches, this implies that weighting implicitly assigns some performance value in instances of method failure. The consequence of our dissuasion from ``missing data-inspired'' approaches is that our work focuses intensively on alternative ways to deal with method failure.

Note that there is an exception to method failure not leading to ``missing'' data. This is when a ``failure'' is recorded but a result could have been obtained; for example, if failure occurs due to excessive runtime but the definition of excessive in the comparison study can safely be judged as shorter than in practice. We refer to this case in Section~\ref{sec:runtime_issues}.

\section{Principles and recommendations for handling method failure} \label{sec3} 
The following section presents guidelines for handling method failure that are more appropriate than discarding data or imputation. Since, as previously mentioned, method failure can only be directly observed after the execution stage, the order of our recommendations aligns with the natural process of investigating these failures post-execution. However, we also highlight key considerations that can be addressed during the planning stage to help reduce both the potential for and unpredictability of failure in the first place. An overview of the recommendations can be found in Table~\ref{checklist}.

Using the terminology of \citet{heinze2024phases}, we distinguish between comparison studies in “early'' and “late'' research phases. Early phase studies investigate if a given method can be ``used with caution'' in settings similar or only slightly different from its original target setting, possibly refining or extending it while also making limited comparisons. Late phase studies, however, focus on realistic comparisons across diverse scenarios to guide real-world users. We make differentiated recommendations when necessary.

\subsection{Practical recommendations for the handling of method failure}\label{sec:recommendations_detailed}

\begin{table}[h]
\caption{\normalsize Checklist of recommendations for handling method failure.}\label{checklist}
\begin{center}
\begin{tabularx}{0.925\textwidth}{|p{0.8\linewidth}|p{0.07\linewidth}|}
   \hline \textbf{Recommended error handling workflow} & Section\\
\hline
\rowcolor{gray!25} 
 \textbf{ 1. Investigate and understand the interplay underlying each instance of method failure} \newline
 \vspace{-0.4cm}\newline \normalsize Interplay may consist of data set characteristics, the method's implementation, the given hardware, and software. & \ref{sec:interplay_method_data} \\
\rowcolor{gray!7}
 \textbf{2. Check for the correct use of the method} \newline \vspace{-0.4cm}\newline \normalsize Distinguish between ``use beyond original scope'' and ``method misuse'' when detecting that a failing method is used differently than intended by its developers.  &
\ref{sec:check_misuse} \\ 
\rowcolor{gray!25}
 \textbf{3. Check if data matches the scope of the study} \newline \vspace{-0.3cm} \newline \normalsize Check if the characteristics of the data associated with failure (and the failure itself) are realistic for the targeted users. & 
\ref{sec:check_scope} \\
\rowcolor{gray!7}
 \textbf{4. Be realistic when modifying method implementations} \newline \vspace{-0.4cm} \newline \normalsize For early-stage research. Restrict yourself to parameter adjustments that are realistic for future targeted users. & \ref{sec:param_modifications}
\\
\rowcolor{gray!25}
 \textbf{5. Consider a fallback strategy} \newline \vspace{-0.4cm}\newline \normalsize For late-stage research. Evaluate the performance of method pipelines of the type ``use method A when it successfully produces an output, use method B otherwise.'' Define the pipelines based on preliminary considerations and avoid selectively reporting the ``best'' pipelines.  & \ref{sec:fallback_strategy} \\

\rowcolor{gray!7}
  \textbf{6. Special case: runtime issues} \newline \vspace{-0.4cm} \newline \normalsize When real users running the affected method only on a single data set can be assumed to successfully obtain a model output, run the method on a preselected random subset of all data sets (selection best made in the planning stage). & \ref{sec:runtime_issues}\\
\rowcolor{gray!25}
 \textbf{7. Report failure proportions, underlying interplays, applied approaches, and share code} \newline \vspace{-0.4cm}\newline \normalsize Provide valuable insights to real users into the methods' usability and enhance a deeper understanding of method failure in general. & \ref{sec:report_proportions}\\
\rowcolor{gray!7}
\textbf{8. Think ahead and pre-register, but remain flexible} \newline \vspace{-0.4cm} \newline \normalsize Conduct a pilot study before pre-registering the handling of method failure. Be aware that unforeseen failures may still occur; therefore, include error-handling mechanisms in your code from the start. & \ref{sec:preregister} \\
\hline
\textbf{Slimmed-down error handling workflow when time constraints prevent recommendations 1-8} & \\ 
\hline
\rowcolor{gray!5}
Report \begin{itemize}[topsep = 0pt]
    \item[(i)] performance results after discarding the data sets for all methods,
    \item[(ii)] performance results after discarding the data sets for failing methods only, and
    \item[(iii)] failure proportions for all methods.
\end{itemize} \vspace{0.2cm}\normalsize Note that this approach does \textbf{not} provide a comparison of unconditional performance.& \ref{sec:slimmed_down}  \\
\hline
    \end{tabularx}
    \end{center}
    \end{table}

\subsubsection{Consider method failure as the result of a complex interplay}\label{sec:interplay_method_data}
When looking beyond the manifestations of method failure summarized in Section~\ref{sec:manifestations}, it becomes evident that these are just the tip of the iceberg. In the end, regardless of how method failure manifests, it is rooted in an individual and complex interplay of interconnected factors, that is 

\begin{center}
    \textbf{certain data set characteristics cause the method to fail in its current implementation (and on the given hardware and software).} 
\end{center}

Besides influencing runtime and memory consumption \citep{zimmermann2020method}, data set characteristics contributing to method failure are highlighted by the observation that a method usually fails for specific data sets, not all of them. Certain data characteristics can complicate or completely prevent required calculations of a method within the maximum number of iterations (non-convergence), time, or memory budgets. Note that this association is addressed in several comparison studies listed in Section~\ref{sec:manifestations}. For instance, the study by \citet{vansmeden2016_10epv} addresses \textit{perfect separation} of the outcome variable by one or more predictor variables as a cause of non-convergence of logistic regression algorithms. Recall also the data set characteristics associated with failure that are reported by \citet{fernandez2014we}, including\textit{``collinearity of data, singular covariance matrices, and equal [predictor values] for all the training [observations], [\dots] discrete predictor variables, classes with low populations, or too few classes''} \citep{fernandez2014we}.

Conversely, the association of method failure with the (implementation of) a given method becomes clear from the usual observation that not all methods fail for the same data sets. That is, methods often vary in their ability to handle specific data set characteristics. For instance, in our illustrative study in Section~\ref{sec:example_OR}, some odds ratio estimation methods are equipped with a way around ``zeros,'' while others fail. 

In addition to the interplay of method and data set characteristics, however, software can also affect the (non-)failure of a method such that results are obtained in one software but not another. Consider, for instance, that \citet{herrmann2021large} report that one of the methods under investigation is completely unusable due to an error in \texttt{R} that occurred using Windows as an operating system, while it does not occur under Ubuntu. The dependency on hardware, on the other hand, becomes clear in how working memory, besides being directly associated with possible memory issues, can also considerably affect runtimes.

Ultimately, the core of method failure--stemming from the interplay of data set, method, software, and hardware--is considerably more informative for addressing method failure than the manifestation of failure itself. We, therefore, recommend that authors of comparison studies investigate and understand the interplay underlying failure for each instance after the execution phase. Ideally, they should identify if particular (and if different) data characteristics cause the failures, leading to evaluations such as ``method A fails when a predictor variable is constant across all observations and when a binary predictor has highly imbalanced classes.'' As a preparatory measure in the planning stage, authors should check user manuals and help pages for any described problems related to certain data structures for all methods under investigation.

\subsubsection{Check for the correct use of the method}\label{sec:check_misuse}
When encountering method failure in the execution stage, authors should first check whether the method was truly developed for the given estimation or prediction task and the type of data (based on the investigations from the previous Section~\ref{sec:interplay_method_data}), and if it is used as intended by the developers. If this is not the case, a distinction should be made between two scenarios. First, a method in the study is used in a slightly different (but not contraindicated) context than originally intended by its developers. We call this ``use beyond the original scope.'' Including this method in the study may be justified when real users commonly use this method ``beyond its original scope,'' however, authors should make a corresponding remark in their manuscript. Second, if using the method in the given context is explicitly contraindicated by the method developers, including it in the study would correspond to method misuse. A method should not be penalized for failing on a data set or a task it was explicitly not intended for. Even if commonly (but probably unknowingly) ``misused'' by real users, this method should be excluded from the study, and the contraindication should be stated clearly in the manuscript. If none of the scenarios apply, authors should resort to Section~\ref{sec:check_scope}.\\
All assessments regarding the use of a given method should be based on the familiarization with all methods from the experiment, which should already take place in the planning stage. This underlines the importance of carefully reading corresponding user manuals and checking for publications from real-world applications in which these methods are commonly used. This aspect should also be checked when planning the data generation in simulation studies and selecting benchmark data for benchmark studies. When in doubt, authors may also consult the developers of the individual methods (and possibly involve them in the study). 

\subsubsection{Check if data matches scope of study}\label{sec:check_scope}

If method failure persists after ruling out an incorrect use in the execution stage, authors should assess whether real-world users are likely to encounter such data (and, therefore, this method failure). This assessment is shaped by the study’s scope and is naturally not purely objective. Importantly, associated considerations should be reported in the study since they carry valuable information and affect the interpretation of the study's results. 

Consider the case of method failure caused by the method's inability to make required calculations due to an ``extreme'' structure of the given simulated data set. For instance, \citet{white2024check} conduct an illustrative simulation study that examines methods for handling missing values in confounders in epidemiological data. They observe failure of an imputation method, and investigations of the corresponding data reveal that it occurs when outcome or exposure are highly imbalanced for individuals with an observed confounding variable. From this information, they conclude that \textit{``the data-generating mechanism is too extreme and should be changed to generate more outcome events''} \citep{white2024check}. \\
If, given the scope of the study, certain data set characteristics are deemed unrealistic for the targeted readers, authors may consider making sensible parameter changes in the DGM to generate data that is more realistic, in line with \citet{white2024check}. 
Similarly, ``too extreme'' data may also affect real data-based studies. If one of the (real) benchmark data sets is considered to be irrelevant to the scope of the study after careful examination, authors may consider removing the data set from the study completely. See Section~\ref{sec:exampleCI} for an illustration.  
However, if the data sets causing failure are deemed relevant to the (simulation or benchmark) study's scope, the researchers conducting the comparison study should---more appropriate than discarding or imputing in most cases---proceed by considering how real-world users are likely to behave in this situation. In particular, they may modify certain parameters of the concerned method(s) (see Section~\ref{sec:param_modifications}) or resort to a fallback (see Section~\ref{sec:fallback_strategy}). 
To be able to assess whether data structures associated with failure are too extreme, certain considerations should already be made in the planning stage. Researchers should be aware of the scope and goal of their study: is it the goal to examine the general behavior of the method(s) (i.e., earlier stages of research) or to provide concrete recommendations for real-world practitioners (later stages of research)? Is the focus on testing the methods under regular or more extreme circumstances? If aiming to provide recommendations to real-world users, what degree of extremity in the data remains realistic for them?

\subsubsection{Be realistic when modifying method implementations} 
\label{sec:param_modifications}
In earlier stages of research, it may make sense to modify a method's implementation entirely (i.e., for all simulation repetitions/benchmark data sets) when encountering method failure during execution. For instance, authors may conclude after execution that the default number of maximum likelihood iterations is generally too low, or that the given optimization algorithm may fail frequently, and that an alternative optimizer is more reliable. This way, authors may modify a greater variety of parameters to see when the method is successful and when it fails. Although this early-stage research is not yet directly aimed at real-world users, authors should remember that it will be used by them later. Therefore, certain practical limitations on parameter modifications should apply to ensure that the method remains feasible for future users.

Also, authors should be careful about increasing the runtimes of affected method(s). Especially raising the number of iterations in optimization algorithms can increase runtime or memory consumption, potentially shifting method failure rather than solving it. A special case of modifying a method's implementation often only arises after the execution stage of the study: investigations following the failure of a method reveal code bugs in the implementation of a method. Especially when their programming experience is high, authors may fix this bug and notify the method developers. Ultimately, the performance assessment of the given method is only meaningful if the authors stress and justify their parameter choices when reporting the study. \\

\subsubsection{Consider a fallback strategy }\label{sec:fallback_strategy}
When a comparison study aims to provide guidance to real-world users (i.e., in later phases of research), it makes sense to mimic their behavior when encountering method failure in the execution stage. How would they proceed when their method of choice failed on their data set? \citet{morris2019using} argue that they would usually not give up but naturally resort to an alternative method, a \textit{fallback}, that successfully provides an output. Therefore, it can make sense to transfer this \textit{fallback strategy} into the context of method comparison studies and evaluate the performance of {\it method pipelines} of the type

\begin{center}
``use method A if it successfully produces an output, use method B otherwise.''
\end{center}
Incorporating fallbacks indirectly yields automatic handling of method failure in case they affect method A but not method B, allowing for the comparison of (absolute and relative) \textit{unconditional} performance when chosen carefully and in a way that there is a suitable fallback for each instance of method failure. Note that when using fallback strategies, we no longer evaluate the ``pure'' performance of method~A. However, it is clear from Section~\ref{sec:caution_discard_impute} that this is unattainable in the first place. \\
The choice of a suitable fallback, again, depends on the scope of the study and the targeted users. Should it serve as a guide for experienced or rather inexperienced users? When targeting inexperienced real-world users, authors may choose a completely different method as fallback. For instance, \citet{zapf2024meta} suggest from the results of their study that \textit{``the approach by Fr{\"o}mke et al (2022) could be used as a fallback strategy in case of non-convergence [of the overall best-performing method] because it always yields results''} \citep{zapf2024meta}, where the approach by \citet{fromke2022semiparametric} is a non-iterative procedure, therefore never affected by non-convergence by design. 
When addressing experienced users, an alternative implementation of the given method (in variation of Section~\ref{sec:param_modifications}) may make a suitable fallback. \\
Many considerations and decisions regarding fallback strategies should already be made in the planning stage of a study. This includes, first and foremost, if using fallbacks is suitable to the given study, and based on this decision and the scope of the study, whether a completely new method or a modified implementation should be used as fallback. Authors should further be aware that the neutrality principle of late phase comparison studies should always apply, that is, authors should avoid putting considerably more effort into only a few methods when choosing fallbacks for them. This becomes clear, for instance, when modifying the implementation of a method, as, e.g., parameter modifications can strongly influence performance \citep{weber2019essential}. In the same context, it is important to avoid selective reporting as it can lead to over-optimistic results; that is, authors should disclose \textit{all} tested pipelines, not just the best ones. It is therefore advisable to reduce the data-driven construction and the complexity of pipelines, which can be achieved by the following considerations. First, authors should decide on a maximum number of fallbacks that can be considered to realistically reflect the behavior of real users (as they are unlikely to try out an indefinite number of methods). Suitable combinations of methods and fallbacks may be preliminarily constructed based on the similarity of ease of use (e.g., a user is unlikely to resort to a fallback that is considerably more difficult to apply than the original method). Alternatively, inspired by \citet{zapf2024meta}, it may make sense to define a default fallback to all instances of method failure if a method has proven to be particularly computationally robust in previous comparison studies. \\
Finally, authors should be aware that instances of all fallbacks failing on a given data set cannot always be avoided, even when using a computationally robust ``default'' fallback. Many fallbacks (and generally many methods) failing on the same data set may indicate that it is particularly ``challenging.'' Authors should then investigate the data set closely to check whether it suits the scope of the study (see Section~\ref{sec:check_scope}) and consider removing it if it is deemed unrealistic for real users. Otherwise, reporting the corresponding characteristics of the data set alongside the methods that successfully produce an output may provide valuable insights to real users.

\subsubsection{Special case: Runtime issues}\label{sec:runtime_issues}
In contrast to non-convergence and memory issues, method failure \textit{is actively declared} by the authors when runtime issues occur in the execution stage. That is, running the methods under investigation on hundreds or thousands of data sets requires time restrictions, which are set by the authors arbitrarily, and without which the method would likely produce a valid output. On the other hand, it can sometimes be safely said that real (targeted) users, running the method on a single data set only, are likely to obtain an output. In these cases, when method failure leads to \textit{missing} rather than \textit{undefined} performance values, it may make sense to run the experiment on a randomly selected subset of all simulation repetitions/benchmark data sets for the method affected by runtime issues (``failure by design''). To ensure that this ``failure'' is independent of data characteristics, it is important to select the subset of data independently of the data sets for which the method has particularly long runtimes. Researchers should therefore think about possible runtime problems and carry out the corresponding random selection of data sets already in the planning stage.

\subsubsection{Report failure proportions, underlying interplays, applied approaches, and share code} \label{sec:report_proportions}

Authors of comparison studies should \textit{report any occurrences and proportions of method failure}. This can serve as an important criterion for real-world users selecting a suitable method. Even the ``best performing'' method is of limited use if the number of settings in which it can be applied without frequent failures is limited. Ideally, the proportion of method failure should be included as an additional performance measure for all considered methods (see \citet{morris2019using}); see \citet{zapf2024meta} for an example.

Additionally, \textit{the interplay of method and data causing failure should be reported}. Adding to the previous paragraph, a corresponding reporting for each method can look like ``method A, with a failure proportion of $X\%$, fails when a predictor variable is constant across all observations and when a binary predictor has highly imbalanced classes.'' In addition to reporting the corresponding failure proportions, readers can learn in \textit{which} settings a method is applicable, and no less importantly, in which it is \textit{not}. Additionally, these insights serve as justification for the chosen handlings of method failure. An example of a corresponding reporting can be found in \citet{hornung2024prediction}, who provide detailed descriptions in the supplement of their paper. \\
Authors of comparison studies should additionally \textit{report their applied handlings} with sufficient detail, that is, any actions resulting from the above-described in-depth investigations and recommendations regarding method failure. This includes removing any misused methods, removing data (or modifying DGMs) that do not match the scope of the study, any changes in implementations, and incorporating fallback strategies. Additionally, authors should justify their choices given the scope of the study (i.e., which users it is aimed at) and the research stage in which it takes place. \\
The complexity of the occurrence and handling of method failure, which becomes clear from the previous sections, emphasizes all the more the value of \textit{sharing code, data, and intermediate results}. Allowing authors of other, ``new'' comparison studies in-depth investigations of the specific instances of method failure in the given study may help them gain a greater understanding of this topic--and enhance fruitful discussions on suitable handlings. 

\subsubsection{Finally: think ahead and pre-register, but remain flexible}\label{sec:preregister}
The previous examples and those that will be outlined in Section~\ref{sec4} make clear that method failure and its cause(s), related to complex dynamics between factors such as method and data set, are often difficult to anticipate. 
While pre-specifying the handling of undefined values has the advantage of reducing analytical uncertainty and its dangers in terms of selective reporting \citep{siepeprereg}, pre-specifying a standard procedure before conducting the study may sometimes be delicate. Making method failure less unexpected (and pre-specifying a procedure for handling method failure less hazardous) can be achieved through pilot experiments. They test whether an experimental design is appropriate for the specific context in which it is applied, and are recommended in several textbooks on \emph{Design and Analysis of Experiments} \citep{cohen1995empirical, dean2017design, montgomery2020design}. For example,  a pilot experiment might use a fraction of the available data sets (e.g., $20\%$) in the comparison study to test whether the specified method implementations produce reasonable results. The remaining $80\%$ is used for the main experiment, which may be refined based on what is learned in the pilot. Pilot studies can also help authors anticipate possible method failures for which they can then pre-specify adequate handlings. Note that it is important to clearly report the results of the pilot study. \\
Despite pre-registration, researchers should be aware that unforeseen issues may appear and require adaptations of the pre-specified strategy (see \citet{lakens2024and} for a discussion of the consequences of deviations from pre-registration on the severity and validity of inferences).  As a result, error-handling mechanisms should be included in the code right from the beginning, including the storage of error messages \citep{white2024check}.

\subsection{Slimmed-down recommendations for the busy researcher (looking to do just enough)}\label{sec:slimmed_down}

For some authors, there may be time constraints that do not allow for any (in-depth) implementations of our recommendations from the previous Section~\ref{sec:recommendations_detailed}, however, authors nevertheless wish to aggregate results over data sets despite the existence of undefined values caused by method failure. While we recommend implementing our in-depth recommendations as comprehensively as possible, we suggest the following slimmed-down analysis and reporting strategy when time and resources are limited. Authors may present performance results after (i) discarding the corresponding data sets \textit{for the failing methods}, (ii) discarding these data sets \textit{for all methods}, and (iii) \textit{report the failure proportions} for all methods, ideally as an additional performance measure. This three-fold analysis and reporting enables more differentiated assessments of the methods' performances and avoids the preference of methods based solely on how data sets associated with method failure are discarded. It also reveals how successful methods perform on data sets for which others fail (and how this influences the method comparison). When the graphical comparison involves boxplots of performance values, authors could indicate method failure (and the corresponding frequency) as scatters of contrasting color above or below the box. For instance, for bias, they could be indicated above the plot (as a value higher than all biases observed in the study), whereas for accuracy, it could be indicated below the plot. 

However, following Section~\ref{sec:caution_discard_impute}, authors should be aware and clearly state in their manuscript that any form of aggregating performance results \textit{despite} existing undefined values does not allow an assessment of \textit{unconditional} performance of all methods under investigation.

\section{Illustrations}\label{sec4}
In the following, we illustrate the guidelines presented in Section~\ref{sec3} through two fictive comparison studies featuring method failure: a statistical simulation study and a benchmark study at the interface between statistics and predictive modeling. 

Note that our focus is exclusively on the handling of method failure and its consequences on the method comparison rather than the method comparison itself. This means that both illustrations represent only a small excerpt (e.g., a few or a single scenario) of what would be a typical publishable study, exclusively focusing on those aspects of the studies that are relevant to the issue of method failure. Importantly, our goal is {\it not} to provide insights into the actual performance of the compared methods.

\subsection{Example 1: Comparing methods for the estimation of odds ratios}\label{sec:example_OR}
\subsubsection{General}\label{sec:OR_intro}

This fictive statistical simulation study deals with the comparison of odds ratio (OR) estimation methods. Outcome and exposure are binary  ($X,Y\in\left\{0,1\right\}$) and the observed number of subjects with $X=i$ and $Y=j$ is denoted as $n_{ij}$. The simulation scenarios are defined by the sample size $n.obs$, the underlying true OR, and the probability of exposure $p_x$. Especially when the sample size is low, the underlying true OR is far from $1$, or $p_x$ is close to $0$ or $1$, some simulated data sets may not contain any subjects for one or more of the four outcome-exposure combinations ($n_{ij}=0$). These ``sampling zeros'' cause calculation issues (e.g., division by $0$) for many OR estimation methods, leading to either returning an error or, depending on the position of the sampling zero in the contingency table, non-meaningful OR estimations of $0$ or ``$\infty$'' for the corresponding data sets. In practice, users often apply the ``Haldane-Anscombe correction'' \citep{haldane1956estimation, anscombe1956estimating} to handle sampling zeros in the OR estimation. This correction consists of adding an offset of $+0.5$ to all four cells in the contingency table to avoid issues in computation. 

\subsubsection{Design of the fictive study}
We vary the underlying \textit{true} OR ($ OR \ \in \{2,5\}$) and exposure probability $p_x$ ($p_x \in \{0.25, 0.5\})$ while keeping the sample size fixed at the relatively low value of $n.obs = 50$. This results in four distinct simulation scenarios (see Table~\ref{tab:prop_samplingzeros} for an overview). For each scenario, we simulate $100\,000$ outcome-exposure data sets. Five OR estimation methods are applied to the simulated data to obtain a point estimate for the true OR. 
``Manual'' corresponds to the manual and straightforward computation $\widehat{OR} = \frac{n_{11} \cdot n_{00}}{n_{01} \cdot n_{10}}$. The remaining four methods, namely ``Fisher,'' ``Midp,'' ``Small'' (all three implemented in \texttt{R} package \texttt{epitools} \citep{R-epitools}), and ``Woolf'' \citep{woolf1955estimating} (implemented in \texttt{R} package \texttt{pairwiseCI} \citep{schaarschmidt2019pairwiseci}) employ more complex modeling strategies.
From the five estimators under investigation, only Small and Woolf can handle sampling zeros internally. 
For each simulation scenario, the estimators are compared regarding bias on the $\log$-scale ($\log$-transformation ensuring symmetry around the value of ``no association'').

Our analysis consists of two parts. First, we compare the performance of the OR estimation methods when \textit{discarding data sets} with sampling zeros for \textit{all} methods versus only for the \textit{failing} methods (``ad hoc approaches''; Section~\ref{sec:OR_adhoc}). Though real users often apply the Haldane-Anscombe correction when encountering sampling zeros, this first part illustrates how the preference of some methods may mainly be driven by the way data sets with sampling zeros are discarded. 

Second, we repeat the analysis implementing the principle of fallback strategies suggested in Section~\ref{sec:fallback_strategy}. This includes important considerations that must be made when choosing suitable fallbacks. For instance, for the manual estimation of the OR (method ``Manual''), only the Haldane-Anscombe correction is applied as a fallback, assuming that a user applying the manual computation in the first place, which is very easy to implement, is unlikely to resort to a method that requires notably more effort. On the other hand, for methods Midp and Fisher for data sets with sampling zeros, we either (i) apply the Haldane-Anscombe correction, (ii) use method Small, or (iii) use method Woolf, assuming that corresponding users may also employ fallbacks that are slightly more difficult to apply. Together with Woolf and Small,  this leads to nine ``methods'' (``pure'' methods or pipelines) being compared in total. The results are presented in Section~\ref{sec:OR_recommended}.

\subsubsection{Occurrence of method failure and ad hoc handlings}\label{sec:OR_adhoc}

Table~\ref{tab:prop_samplingzeros} displays the proportion of the $100 \, 000$ simulated data sets containing at least one sampling zero across the four simulation scenarios. As expected, this proportion increases with an increasing true OR$>1$, while decreasing when exposure $X$ is more balanced. 

\begin{table}[t]
\centering
\caption{Overview of the proportion of the $100 \, 000$ simulated data sets (fixed sample size $n.obs = 50$) with sampling zeros across the simulation scenarios. $p_x$: exposure probability.}
\normalsize
\label{tab:prop_samplingzeros}
\begin{tabular}{c c  c c}
\toprule
Simulation scenario & \textit{true} OR & $p_x$ & Sampling zero proportion \\
\midrule
1 &  2  &$0.25$ &1.28\%\\
%2 &  3  &$0.25$& 3.96\%\\
%3 &  4  &$0.25$& 7.8\%\\
2 &  5  &$0.25$& 12.0\%\\
3 &  2  &$0.5$& 0.01\%\\
%6 &  3  &$0.5$& 0.11\%\\
%7 & 4  &$0.5$& 0.50\%\\
 4 & 5  &$0.5$& 1.27\%\\
\bottomrule
\end{tabular}
\end{table}

\begin{figure}[t]
\centering
\includegraphics[width = \textwidth]{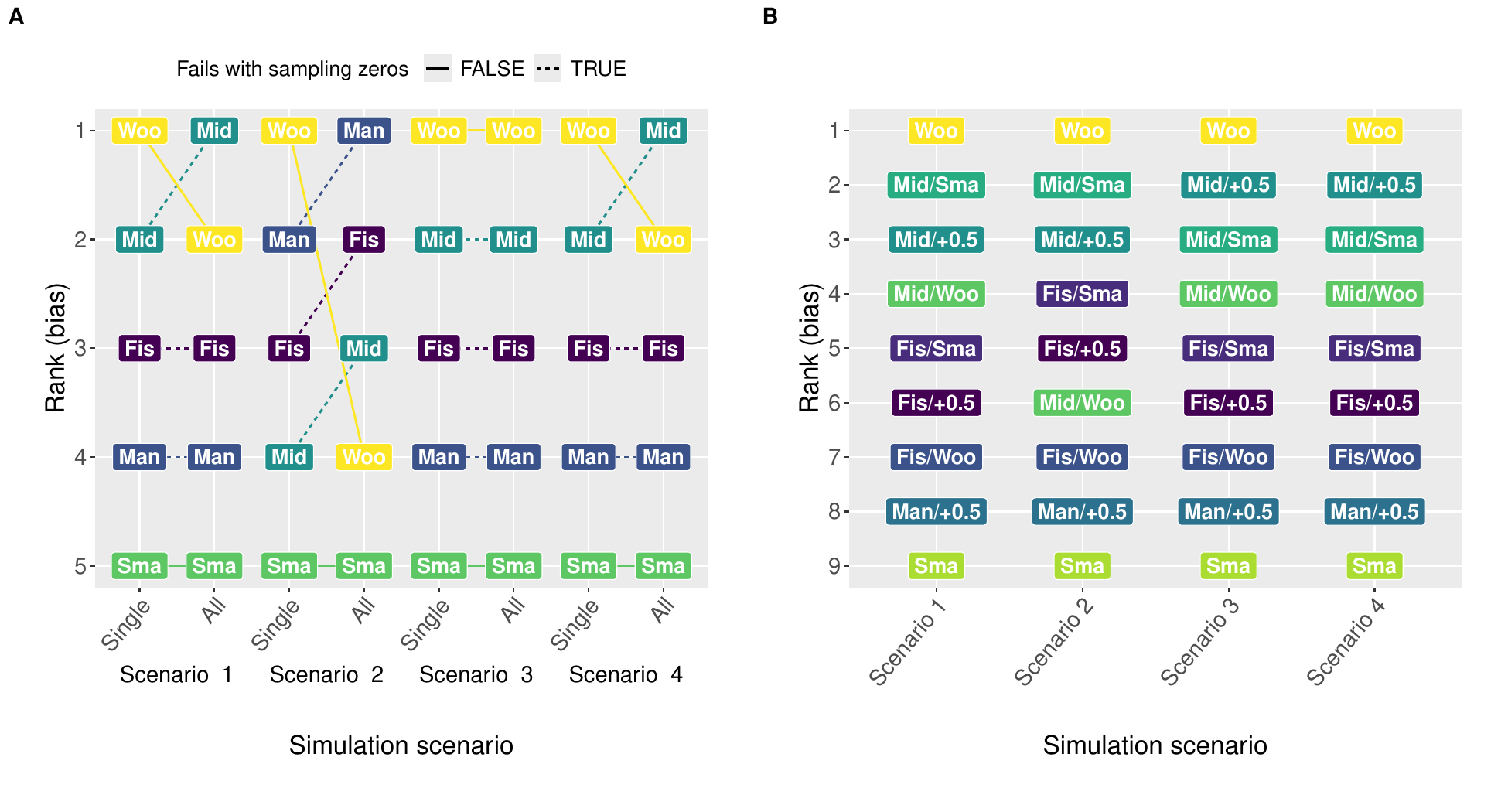}
 \caption{Performance ranks of the OR estimation methods across the eight simulation scenarios based on empirical bias on the $\log$-scale, where a rank of 1 means that the method has the lowest empirical bias in the given scenario. \textbf{(A)} Ranks are compared between \textbf{discarding data sets} for the individual failing methods only (\textbf{``Single''} on the $x$-axis) versus for all methods (\textbf{``All''} on the $x$-axis). \textbf{(B)} The ranks are compared after \textbf{applying fallback strategies} to methods Midp, Fisher, and Manual. For Midp and Small, the fallbacks Small, Woolf, and the Haldane-Anscombe correction (``$+0.5$'') are used. For Manual, only the Haldane-Anscombe correction is used as fallback. Small and Woolf can deal with sampling zeros internally.\\
 Abbreviation of methods: ``Fis'': Fisher; ``Mid'': Midp``; ``Sma'': Small; ``Woo'': Woolf; ``Man'': Manual (i.e., manual estimation of the OR). \textbf{(B)}: The combination of original method $m_1$ and fallback $m_2$ is marked by ``$m_1\slash m_2$.''}\label{fig:OR_performanceranks}
\end{figure}

\textbf{Ad hoc approaches} \ 
In the first analysis, Figure~\ref{fig:OR_performanceranks}(A) demonstrates that the way of discarding data sets with sampling zeros can substantially affect the results of the method comparison, with discrepancies regarding performance ranks already observable at a sampling zero proportion of just over $1\%$. This impact becomes more pronounced as the proportion of sampling zeros increases, making a difference in the observed performance by up to three ranks. In particular, it can affect which method is perceived as best-performing. Note that a more detailed investigation of Woolf's performance shows that for the data sets \textit{without} sampling zeros, Woolf slightly underestimates the true OR, while for the few data sets \textit{with} sampling zeros, Woolf notably overestimates it. This leads to top performance when all data sets are included in its performance assessment (while discarding data sets with sampling zeros for the failing methods).

\subsubsection{Recommended handling: using fallbacks
}\label{sec:OR_recommended}
Woolf's overall good performance across all scenarios is underpinned when using fallbacks for all methods affected by method failure. Furthermore, the comparison between Woolf and Midp, another method performing well in most scenarios, becomes clearer through the use of fallbacks. While they often alternate in rank (and especially between rank 1 and 2) when removing data sets (Figure~\ref{fig:OR_performanceranks}~(A)), their performance comparison is constant across all scenarios when a fallback is applied to Midp (Figure~\ref{fig:OR_performanceranks}~(B)). Since this applies to both fallbacks, the Haldane-Anscombe correction and Small, it is debatable which fallback is preferable. Both fallbacks for Midp frequently switch ranking positions across the performance measures and simulation scenarios, however, fallback Small might be even easier to implement than the Haldane-Anscombe correction, requiring only a simple code modification from Midp. 

\subsection{Example 2: Comparing methods for computing confidence intervals for the generalization error}\label{sec:exampleCI}

\subsubsection{General}
The fictive study aims to compare the performance of two methods for constructing confidence intervals (CIs) for the generalization performance of a prediction model based on resampling-based estimates, such as cross-validation. The ``naive'' method ``$N$'' for constructing the CIs ignores the dependence between the generalization performances in the resampling repetitions, while method ``$C$'' corrects for it. The performance of both methods is assessed in terms of \emph{coverage}. Note that a comprehensive benchmark study of methods for constructing CIs for the generalization error is performed by \citet{schulz2024constructing}.

\subsubsection{Design of the fictive study}
Classification trees \citep{breiman1984cart} are considered as models for predictive modeling and the data set provided by the SAPA project \citep{condon2017sapa}, further processed according to \citet{klau2023comparing}, is used as the benchmark data set. The response variable is binary. The generalization performance is quantified using the Area Under the Curve (AUC).
The AUC is estimated using $15$ repetitions of repeated subsampling ($i=1,\ldots,15$) with a 4:1 ratio (i.e., the data is randomly split $15$ times into $80\%$ on which the model is trained and $20\%$ on which the AUC is obtained \citep{Hastie2009}).  

Method $C$ suggested by \citet{nadeau1999inference} computes the CI for the generalization AUC as
\begin{align}\label{eq.ci_correct}
    \Big[ \ \overline{\widehat{AUC}} - t_{0.975,14} \cdot  \sqrt{\Big(\frac{1}{15}+c\Big)\cdot S^{2}_{\widehat{AUC}_{i}}}, \quad \overline{\widehat{AUC}} + t_{0.975,14} \cdot \sqrt{ \Big( \frac{1}{15} + c \Big) \cdot S^{2}_{\widehat{AUC}_{i}}} \ \Big],
\end{align}
where $\overline{\widehat{AUC}}$ represents the mean AUC estimate and $S^{2}_{\widehat{AUC}_{i}}$ the sample variance over the $15$ subsampling repetitions. $t_{0.975,14}$ is the corresponding quantile of the $t$-distribution. A fixed correction term $c>0$ incorporates the correlation structure between the subsampling repetitions for method $C$ (see \citet{nadeau1999inference} for the construction of c). Method $N$, on the other hand, sets $c=0$ in the above formula. 

\begin{figure}[t]
\centering
\includegraphics[scale = 0.65]{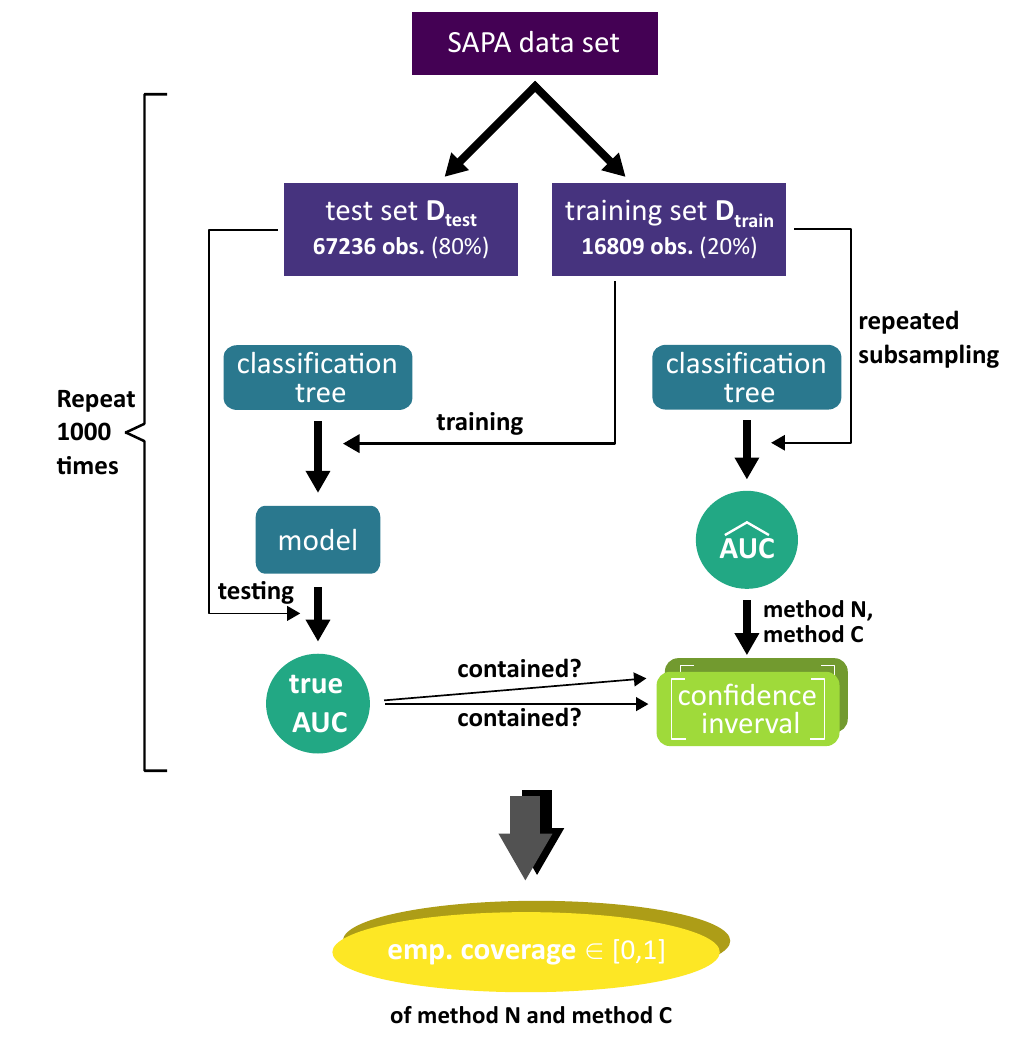}
 \caption{Design of the fictive study to obtain the empirical coverage of methods $C$ and $N$. An approximation of the  ``true AUC'' is obtained based on the very large $D_{test}$.}\label{fig_exCI_design}
\end{figure}

To estimate the coverages of methods $N$ and $C$, the following procedure is repeated for $1000$ iterations (see Figure~\ref{fig_exCI_design}). The SAPA data set is randomly split into a large test set \textit{$D_{test}$} ($80\%$ of observations) and a training set \textit{$D_{train}$} (20\% of observations). The \textit{true} AUC is approximated by applying the trained model to $D_{test}$. The corresponding \textit{estimated} AUC, on the other hand, is obtained via repeated subsampling on $D_{train}$. Confidence intervals for the true AUC are constructed according to Equation~ 
(\ref{eq.ci_correct}) with corresponding values of $c$ for methods $N$ and $C$.
Finally, the estimated coverages of methods $N$ and $C$ are the proportions of the $1000$ iterations where the respective confidence intervals cover the approximated true AUC.

The study is conducted in \texttt{R} using package \texttt{mlr3}, method $N$ is implemented using function \texttt{t.test()} from package \texttt{base} and for method $C$, $\overline{\widehat{AUC}}$ and $S^{2}_{\widehat{AUC}_{i}}$ are obtained using functions \texttt{mean()} and \texttt{sd()} (from the same package).

\vspace{0.3cm}

\subsubsection{Occurrence of method failure and ad hoc handlings}

For method $N$, the implementation of the above-described procedure returns the following error message for 300 of the $1000$ iterations in \texttt{R}: 
\begin{center}
    ``\texttt{Error in t.test.default: data are essentially constant.}'' 
\end{center}

Hence, the confidence intervals cannot be constructed for method $N$ in these iterations, making it impossible to assess whether they cover the true AUC. Method $C$ is not affected by these errors. At first glance, the error message gives little indication of how to resolve it. In the following, we first apply three ad hoc approaches to handle these errors. Then, by following our in-depth guidelines from Section~\ref{sec3}, we demonstrate how these ad hoc approaches prove to be inadequate, and illustrate more suitable ways to handle method failure in this example. \\

\textbf{Ad hoc approaches} \ One could be inclined to apply one of the following ad hoc approaches: 
\begin{itemize}[noitemsep]
    \item[1.] For method $N$, ignore all data sets in which it fails (method $C$ is unaffected). 
    \item[2.] For both methods, ignore those iterations in which $N$ fails. 
    \item[3.] For method $N$, set ``undefined'' confidence intervals to ``not covering the true AUC.'' 
    \end{itemize}
    
Note that the intuition to approach~3, which can be viewed as a form of imputation, is that a CI that does not exist cannot cover any value. It thus induces a negative correlation between the proportion of undefined confidence intervals and the coverage of method $N$. 

\begin{table}[h]
\caption{\normalsize Empirical coverages of methods $N$ and $C$ resulting for (a) the three ``ad hoc'' approaches and (b) setting undefined CIs to zero-width CIs. }\label{emp.cov_adhoc}
\begin{center}

   % \resizebox{\textwidth}{!}{%
   \renewcommand{\arraystretch}{1} % Default value: 1
    \begin{tabularx}{0.98\textwidth}{p{0.1\linewidth}Y{0.17\linewidth}Y{0.18\linewidth}Y{0.19\linewidth}Y{0.0001\linewidth}Y{0.18\linewidth}}

    \toprule
   \multicolumn{1}{c}{} &   \multicolumn{3}{c}{(a) Ad hoc approaches: NA-CIs are} & &\multicolumn{1}{c}{(b) NA-CIs are}\\

     \addlinespace[0.2cm]
     \cline{2-4} \cline{6-6}%\cline{8-11}

    \addlinespace[0.2cm]
&  discarded for\newline method $N$ only &  discarded for\newline methods $N$ and $C$& set to ``not covering  \newline true AUC'' & & set to zero-width \newline interval\\
    \midrule

Method $N$  &   $0.54$ & $0.54$ & $0.38$ & &$0.68$\\
Method $C$ &    $1$  & $1$  & $1$ & &$1$ \\

 \bottomrule
    \end{tabularx}

    \end{center}
        \end{table}

The coverages resulting from the ad hoc approaches are displayed in Table~\ref{emp.cov_adhoc}~(a). While the coverage of method $C$ is constant across all three ad hoc approaches, method $N$, while too liberal across all approaches, is even more liberal using the ad hoc imputation approach.  \\

\subsubsection{Recommended handling}
We now closely inspect the modeling processes underlying the study to find the root cause of the error message and to adapt the handling of method failure accordingly. \
Investigations show that the errors in method $N$ occur when the estimated AUCs are equal across all $15$ subsampling repetitions, resulting in a sample variance $S^{2}_{\widehat{AUC}{i}}$ of 0. This zero variance in the denominator of the $t$-statistic causes an error in function \texttt{t.test()}. Focusing on the confidence interval, however, a sample variance of 0 simply implies a zero-width CI $\left\{\overline{\widehat{AUC}}\right\}$. Therefore, method~$N$ not yielding a CI (and returning an error) is a technical artifact of its implementation. A straightforward solution would be to modify the implementation of method $N$, using \texttt{mean()} and \texttt{sd()} instead of \texttt{t.test()}, so that it computes the zero-width CI, which can enter the calculation of coverage as any other CI (see Table~\ref{emp.cov_adhoc}~(b) for the resulting coverage). \\
However, upon yet further investigations, it becomes clear that a sample variance $S^{2}_{\widehat{AUC}{i}} = 0$ always coincides with all 15 estimated (and the approximate true) AUC being $0.5$. Ultimately, this worst possible performance of the classification tree (``random prediction'') occurs when the classification tree does not perform any splitting. This is related to the characteristics of the data set. For example, the problem disappears if we include additional predictors: splitting then occurs, leading to an AUC higher than $0.5$. 

This illustrates well that the further the investigations are taken, the apparent adequate handling of method failure can change notably. In this case, based on the information gained, adequate handling depends on the aim of the study and requires further considerations such as the following. A prediction problem where splitting often does not occur can be viewed as an edge case. If the study's scope includes such edge cases, one could attempt to suitably modify the parameters of the classification tree to favor splitting. If, however, the study aims to compare both methods in regular scenarios, one might consider adding additional predictors to favor splitting more strongly, resort to an alternative prediction model (e.g., logistic regression), an alternative measure of generalization performance (e.g., accuracy), or use a different benchmark data set. 

\section{Discussion}
Method failure in comparison studies, often manifesting as issues regarding calculation (e.g., non-convergence), runtime, or memory issues, complicates the desired comparison of \textit{unconditional} performance between the methods under investigation, i.e., across all simulated or benchmark data sets. Our informal literature review and illustrations both make clear that authors should not blindly rely on the initially perceived manifestation of failure but rather make extensive investigations of the underlying factors causing the failure, which often guides the choice of the handling of the resulting undefined values. 

Even if we argue for decisions on a case-by-case basis after careful inspection of the failures, we show and illustrate that some general principles hold in most situations. In particular, method failure usually leads to non-existing values instead of just ``missing,” i.e., existent but unobserved values. This means imputation, a common but missing-data-inspired approach to handling method failure, is usually inadequate. The same applies to the popular and seemingly logical approach of discarding data associated with failure (either for all or the failing methods), which almost never enables the comparison of unconditional performance. In this work, we provide alternative approaches that might be more appropriate to most occurrences of method failure. 

An approach that in our view deserves more attention than it receives currently in the literature consists of evaluating whole \lq\lq pipelines'' of methods, including one or several fallback strategies in case of method failure. This approach reflects the typical behavior of most users of methods in practice while enabling the comparison of unconditional performance if constructed carefully: if one method does not work, they resort to an alternative one. Future work is required to collect experiences and define standards regarding such pipelines and their evaluation.  Above all, transparent reporting of the frequency of failure and of the corresponding handling is always recommended. 
 
A limitation of our recommendations is that, beyond obvious cases, they typically require both high time resources and thorough expertise in the underlying modelling processes and methods. Researchers conducting comparison studies may not always have these resources to the full extent. This particularly applies to a nuanced examination of the interactions between methods and data sets leading to method failure in different instances. We understand that this bottleneck may impair the full implementation of our recommendations in practice, hence our attempt to provide a slimmed-down version of the recommendations. When time resources and expertise are restricted, transparent reporting is all the more important. Furthermore, this underlines the necessity for openly sharing code, data, and ideally intermediate results, as this allows other researchers to explore the observed instances of method failure in greater detail. 
Also, our viewpoint on method failure as leading to ``undefined” performance values does not always hold unrestrictedly, which can complicate the implementation of our recommendations.  However, with method failure being a complex matter in and of itself, this underscores the importance of a case-by-case evaluation, as described above. 

In this article, method failure has been treated as black-or-white. However, in practice, whether or not a method has failed is itself a value of judgment. One such example is near-separation \citep{vansmeden2016_10epv}, which does not necessarily make itself obvious, hence the existence of dedicated methods to detect it \citep{Mansournia2017}. Such gray-area failures add further complexities to benchmark and simulation studies: overall performance may depend on the specific criteria used to judge failure.  

Partial failure of a method is also possible. For example, in a simulation study, a method might return a valid point estimate but not its standard error, or a point estimate for one estimand but not another. Many of the principles discussed in this article apply in such situations, for example, the principle to avoid imputing estimates where a method has failed, and the notion of a pipeline (when the standard error estimator fails, revert to a backup estimator if available). There are, however, further subtleties. For example, it may be inadvisable for the fallback approach to keep the point estimate from method~A and fallback on the standard error estimated from method~B.

To sum up, we believe there is not ``one” right way to handle method failure in all situations. Instead, it is important to carefully choose an approach depending on the context, and then report, explain (and ideally discuss) the decisions made. With our work, we hope to contribute meaningful ideas on how to think about method failure. They lead to helpful strategies for dealing with it, as outlined as part of our recommendations. However, we also believe that method failure is only one of many issues related to comparison studies that deserve much more time, effort, and methodological strength towards more reliable evidence on the behavior of methods in the context of a \lq\lq replication crisis in methodological research'' \citep{boulesteix2020replication}. We need more well-designed neutral comparison studies \citep{boulesteix2017towards}---late phase studies according to the terminology of \citet{heinze2024phases}. In such studies, no time is spent on the development of new methods. This leaves time for implementing our recommendations and diving into the technical details of the methods under investigation.

\section*{Funding Information}

{\label{974317}}
This work is supported in part by funds from the German Research Foundation (DFG: BO3139/7 and BO3139/9-1). TPM was funded by the UK Medical Research Council (grant MC\_UU\_00004/09).

\section*{Acknowledgements}
The authors thank Julian Lange for useful literature input and Clemens Kreutz for valuable comments. The authors further thank Savanna Ratky for language corrections and Luzia Han{\ss}um for support with formatting the manuscript. 
{\label{749861}}

\section*{Conflicting interests}
The authors have declared no conflicts of interest for this article.

\section*{Data availability statement}The code to reproduce all results from Section~\ref{sec4} are available on
\href{https://github.com/chillemille/ProjMPV}{\color{gray}{Github}}.

\FloatBarrier
\bibliographystyle{apalike}
\bibliography{reference}

\newpage
\beginsupplement
%\FloatBarrier
%\bibliographystyle{apalike}
%\bibliography{reference}
\end{document}